\documentclass[preprint,preprintnumbers,amsmath,amssymb,12pt]{revtex4}
\usepackage{graphicx}
\usepackage{epsfig}
\usepackage{bm}
\usepackage{amsfonts}
\usepackage{subfigure}
\usepackage{multirow}
\usepackage{color}
\usepackage{float}
\usepackage{xcolor, soul}
\usepackage{array}
\newcolumntype{C}[1]{>{\centering\arraybackslash}p{#1}}
\graphicspath{{./figs/}}
\sethlcolor{green}

\usepackage{hyperref}
\hypersetup{colorlinks=True,citecolor=blue}
\begin{document}

\title{\textbf{GUP-modified $\alpha$-attractor inflation in light of ACT observations}}
\title{\textbf{$\alpha$-attractor inflation modified by GUP in light of ACT observations}}

\author{Hana Heidarian\footnote{hanna.heidarian@uok.ac.ir}, Milad Solbi\footnote{miladsolbi@gmail.com}, Soma Heydari\footnote{s.heydari@uok.ac.ir}, and Kayoomars Karami\footnote{kkarami@uok.ac.ir}}
\affiliation{\small{Department of Physics, University of Kurdistan, Pasdaran Street, P.O. Box 66177-15175, Sanandaj, Iran}}

\date{\today}

\begin{abstract}
Here, the $\alpha$-attractor inflation is investigated within a framework incorporating a minimal measurable length, as implemented by the Generalized Uncertainty Principle (GUP). The GUP modifications to the Friedmann equations and cosmological perturbation parameters are employed to assess the model observational viability against the Atacama Cosmology Telescope (ACT) data. Our results indicate that in the $r-n_s$ plane, the predictions of the standard $\alpha$-attractor model ($\beta=0$) lies near the $2\sigma$ boundary of joint observations. More interestingly enough is that in the presence of GUP effect, the predictions of the model for the GUP parameter $\beta \gtrsim O(10^{13})$ shifts into the $68\%$ CL interval. This value for $\beta$ is in well agreement with upper bounds on the GUP parameter deduced from cosmological analysis as well as quantum and gravitational experiments.
\end{abstract}


\maketitle

\section{Introduction}

The theory of cosmic inflation is now widely accepted as a fundamental component of the standard cosmological model.
It was originally proposed to address several critical shortcomings inherent in the Hot Big Bang theory, such as the horizon, flatness, and primordial monopole problems \cite{Guth:1981,Linde:1982,Starobinsky:1980}.
Beyond resolving classical puzzles, the inflationary era provides a compelling mechanism for generating primordial density perturbations.
These perturbations are originated from quantum fluctuations of the scalar inflaton field \cite{Mukhanov:1981,Hawking:1982,Bardeen:1983,Guth:1982pi,Starobinsky:1982ee}.
After inflation ends, these fluctuations seed the formation of large scale structures and imprint the observed anisotropies in the Cosmic Microwave Background (CMB) radiation.

Over the past decades, measurements of CMB anisotropies have been progressively refined through a series of observational missions. The COBE satellite first detected anisotropies on large angular scales.
This was followed by WMAP~\cite{WMAP:2013}, and more recently, the Planck satellite~\cite{Planck:2018inflation}, which provided high resolution, full sky measurements with unprecedented precision.
Complementary ground based experiments, such as the Atacama Cosmology Telescope (ACT)~\cite{ACT:DR6params,ACT:DR6models} and the South Pole Telescope (SPT)~\cite{SPT:2022filters}, have improved these observations.

Among various inflationary frameworks, the $\alpha$-attractor models have received significant attention \cite{Kallosh:2013a,Kallosh:2013b,Ferrara:2013,Ellis:2013,Kallosh:2014,Rezazadeh:2022,Teimoori:2021,Carrasco:2015uma}.
These models are well motivated by supergravity and string theory \cite{Kallosh:2013b,Ferrara:2013,Kallosh:2014,Carrasco:2015uma}.
Moreover, their predictions for key observables such as the scalar spectral index $n_s$ and the tensor-to-scalar ratio $r$ were in excellent agreement with earlier observational data \cite{Planck:2018inflation,bk:18}.
These features have made $\alpha$-attractor models particularly attractive in the context of early universe cosmology.
However, recent observational data from ACT Data Release 6 (ACT DR6), when combined with Planck, DESI BAO, and BICEP/Keck, suggest a slightly higher value for the scalar spectral index $n_s$ compared to that inferred from earlier data, namely $n_s = 0.974 \pm 0.003$~\cite{ACT:DR6params,ACT:DR6models}. Given these updated constraints, it is necessary to reassess well motivated inflationary scenarios \cite{Kallosh:2025act,Aoki:2025,Berera:2025,Dioguardi:2025,Raidal:2025,Salvio:2025,Antoniadis:2025,Kim:2025,Drees:2025,Haque:2025,Peng:2025bws,Liu:2025qca,Gialamas:2025ofz,He:2025bli,Wolf:2025ecy,Dioguardi:2025mpp}.
Within this context, the predictions of $\alpha$-attractor models for $N = 60$ now lie close to the edge of the $2\sigma$ region, and are excluded for lower allowed values of the $e$-folds number~\cite{ACT:DR6params,ACT:DR6models}.
This discrepancy motivates refinements to the theoretical predictions of $\alpha$-attractor models and encourages the exploration of additional physical effects that may reconcile these models with the latest cosmological observations.

The possible effect of quantum gravity considerations is an interesting area for theoretical exploration.
Specifically, the existence of a fundamental minimal length, which is expected to be of the order of the Planck length $l_{\rm p}$, may have an significant impact.
The concept of minimal measurable length arises in various approaches to quantum gravity, including string theory~\cite{Amati:1989,Gross:1988,Konishi:1990}, loop quantum gravity~\cite{Rovelli:1995,Ashtekar:1997}, asymptotically safe gravity~\cite{Lauscher:2005,Reuter:2006}, and non-commutative geometry~\cite{Seiberg:1999,Connes:2000}.

The presence of a minimal length implies that the standard Heisenberg Uncertainty Principle (HUP) extended to a Generalized Uncertainty Principle (GUP), a concept first introduced in the pioneering work \cite{Kempf:1995}.
For a comprehensive overview, see the review \cite{Bosso:2023aht} and references therein.
Various GUP models and their phenomenological implications have been extensively explored in the literature
\cite{Luo:2025,Bitaj:2024,EPL:2023,Du:2022,Feng:2022,Luciano:2021,Mohammadi:2017,Medved:2004,Ali:2012,Awad:2014,Zhu:2009,Das:2008,Gialamas:2024ivq,Segreto:2024vtu,Banerjee:2022slh,Feng:2020ams,Bambi:2007ty,Barca:2023epu}.
A widely studied GUP model incorporates a quadratic correction in momentum, which is given by
\begin{equation}
\Delta x \geq \frac{\hbar}{\Delta p} \left[1 +  \frac{\beta\,l_{\rm p}^2}{\hbar^2} \left( \Delta p\right)^2 \right],
\end{equation}
where $\beta$ is a dimensionless GUP parameter~\cite{Bitaj:2024,EPL:2023,Du:2022,Luciano:2021,Feng:2022}.
It is worth noting that other GUP formulations have also been investigated, which include formulations with higher-order momentum terms or different functional forms~\cite{Luo:2025,Du:2022,Feng:2022,Kempf:1995}.
These GUP frameworks lead to a modification of the fundamental canonical commutation relations.
Physical observables that depend on these commutation relations or the density of states are expected to receive GUP corrections.
In the context of cosmology, a key implication is the modification of the momentum, and consequently the wavenumber $k$ of quantum fields \cite{Bitaj:2024,EPL:2023,Du:2022,Luciano:2021,Feng:2022}.
This is particularly relevant for the quantum fluctuations which are generated during inflation.

The influence of a minimal length on inflationary cosmology can be determined through several distinct mechanisms.
One direct consequence is the modification of the dispersion relation for quantum fields.
Equivalently, a deformation in the relationship between momentum and wavenumber occurs.
In many GUP scenarios, the physical wavenumber $K$ relevant to perturbations is no longer simply proportional to the comoving wavenumber $k$.
Instead, it becomes a function as $K(k)$.
In this regard, a common form of such a modification arises from GUP models that alter the momentum operator as $P^i = p^i(1 + \beta\,l_{\rm p}^2\,p^2)$, which leads to a modified wavenumber $K^i = k^i(1 + \beta\,l_{\rm p}^2\,k^2)$ ~\cite{EPL:2023,Bitaj:2024}. Note that here we have set $\hbar=1$ in the relation $p=\hbar k$.
Considering such a modification to $K$ can directly impact the predicted inflationary observational quantities.
Specifically, the definitions of the scalar and tensor spectral indexes are modified as follows \cite{EPL:2023,Bitaj:2024}
\begin{equation}\label{eq:nsnt}
n_s \simeq 1 + (1 + \beta\,l_{\rm p}^2\,k^2)~\left( \frac{d\ln \mathcal{P}_s}{d\ln k} \right)_{\text{cl}}, \quad
n_t \simeq (1 +\beta\,l_{\rm p}^2\,k^2)~\left(\frac{d\ln \mathcal{P}_t}{d\ln k}\right)_{\text{cl}},
\end{equation}
where $\mathcal{P}_s(k)$ and $\mathcal{P}_t(k)$ denote the scalar and tensor primordial power spectra, respectively.
Here, the subscript "cl" refers to the classical standard calculation without GUP corrections.
These modifications to $n_s$ and $r$ may potentially affect the predictions of inflationary models in comparison with the observational data.

In addition, taking the GUP into consideration, can also influence inflationary models by modifying the Friedmann equations \cite{Bitaj:2024,Mohammadi:2017,Awad:2014}.
In the thermodynamical interpretation of gravity, the apparent horizon entropy follows the area law, $S = A/(4\,l_{\rm p}^2)$, which is corrected under the
GUP consideration by the parameter $\beta$.
These quantum corrections lead to modified Friedmann equations \cite{Zhu:2009,Awad:2014,Bitaj:2024}, thereby modifying the Hubble rate and the dynamics of the inflaton field. As a result, the slow-roll parameters and consequently the predictions for $n_s$ and $r$ are modified.

Given the increasing precision of CMB observations and the theoretical motivation for the minimal length effects, it is timely to investigate whether GUP corrections can improve the agreement between the $\alpha$-attractor models and the latest observational data. If such corrections shift predicted values of $n_s$ and $r$ closer to the observed ranges, this could not only enhance the viability of these inflationary models, but also provide indirect evidence for quantum gravitational effects at inflationary scales. Additionally, this work draws a comparison between the model predictions and observational data which may yield new constraints on the GUP parameter $\beta$.

The paper is organized as follows. In Section~\ref{sec2}, we review the $\alpha$-attractor inflation modified by GUP. Section~\ref{sec:observational_viability} is devoted to analyzing the evolution of the background parameters under the minimal length effects and comparing them with the standard inflationary scenario. We then assess the observational viability of the model. Finally, the main conclusions are presented in Section~\ref{conc}.
\section{\texorpdfstring{$\alpha$-attractor inflation modified by GUP}{alpha-attractor inflation modified by GUP}}\label{sec2}
Here, the action of standard inflationary model is introduced as follows \cite{Guth:1981,Linde:1982}
\begin{equation}\label{eq:action}
S=\int{\rm d}^{4}x \sqrt{-g} \left[\frac{M_{\rm p}^2}{2}R+ X - V(\phi) \right],
\end{equation}
where $g$ and $R$ are the determinant of the metric tensor $g_{\mu\nu}$ and the Ricci scalar, respectively. In addition, $M_{\rm p} \equiv 1 / {\sqrt{8\pi G}}$ indicates the reduced Planck mass, $X\equiv\frac{1}{2}g^{\mu\nu}~\partial_\mu \phi \partial_\nu \phi$ and $V(\phi)$ denote the kinetic energy term and the scalar field potential, respectively. The spatially flat Friedmann-Robertson-Walker (FRW) metric is given by
\begin{equation}
\label{eq:FRW}
{\rm d} {s^2} = {\rm d} {t^2} - {a^2}(t)\left( {{\rm d} {x^2} + {\rm d} {y^2} + {\rm d} {z^2}} \right) ,
\end{equation}
where $a(t)$ is the scale factor and $t$ is the cosmic time. Hence, for a homogeneous scalar field $\phi(t)$, the kinetic term simplifies to $X=\dot{\phi}^2/2$.

The Friedmann equations are obtained by taking the variation of the action (\ref{eq:action}) with respect to the metric (\ref{eq:FRW}) as follows
\begin{align}
&H^{2} =\dfrac{1}{3M_{\rm p}^2} \rho_{\phi} , \label{eq: FR-eqn1} \\
&\dot{H} = -\dfrac{1}{2M_{\rm p}^2} (\rho _{\phi} +p_{\phi} )  ,
\label{eq:FR-eqn2}
\end{align}
where $H \equiv \dot{a}/a$ stands for the Hubble parameter, and the dot represents the derivative with respect to cosmic time $t$. Furthermore, the energy density $\rho_{\phi}$ and pressure $p_{\phi}$ of the scalar field in the standard inflationary model are given by
\begin{equation}
\rho _\phi =\frac{1}{2}\dot{\phi }^{2} + V(\phi) \label{eq:rho},
\end{equation}
\begin{equation}
{p_\phi } = \frac{1}{2}\dot{\phi }^{2}- V(\phi) \label{eq:Lp}.
\end{equation}
It has been proven that the Friedmann equations are modified in the presence of a minimal length \cite{Bitaj:2024,Mohammadi:2017,Awad:2014}.
The modified Friedmann equations can be derived from thermodynamical considerations.
This approach is based on applying the first law of thermodynamics, $dE = TdS$, to the apparent horizon of the universe.
The GUP introduces a correction to the standard Bekenstein-Hawking entropy and consequently the standard entropy-area relation is modified.
This GUP-corrected entropy, when used in the first law, results in modified Friedmann equations.
For a detailed, step-by-step derivation of this procedure, we refer the reader to the work by Awad and Ali \cite{Awad:2014}.
The modified Friedmann equations are given as

\begin{align}
&\frac{H^{2}}{2}+\frac{1}{3\,\beta\,l_{\rm p}^{2}}\,\Bigg[1-\bigg(1-\beta\,l_{\rm p}^{2}\,H^{2}\bigg)^{\frac{3}{2}}\Bigg]=\frac{1}{3M_{\rm p}^2}\,\rho_{\phi}, \label{eq:FR1-ML}\\
&\dot{H}\Bigg(\frac{\beta\,l_{\rm p}^{2}}{2}\Bigg)\Bigg(\frac{H^{2}}{1-\sqrt{1-\beta\,l_{\rm p}^{2}\,H^{2}}}\Bigg)=-\frac{1}{2M_{\rm p}^2}\,(\rho_{\phi}+p_{\phi}). \label{eq:FR2-ML}
\end{align}
Moreover, the equation of motion for the scalar field $\phi$ is derived from the energy conservation law, $\dot{\rho}_\phi + 3H(\rho_\phi + p_\phi) = 0$. By substituting Eqs. (\ref{eq:rho}) and (\ref{eq:Lp}) in the conservation law equation the equation of motion of the scalar field  is derived as follows

\begin{equation}\label{eq:KG}
\ddot{\phi }+3H\dot{\phi }+V'(\phi )=0,
\end{equation}
where $({}')$ denotes the derivative with respect to $\phi$.
It is crucial to note that while the form of this equation is standard, the GUP effect is included implicitly.
The Hubble parameter, $H$, which appears in Eq. (\ref{eq:KG}) is determined by the GUP-modified Friedmann equations.
Therefore, the evolution of the scalar field $\phi$ is inherently sensitive to the minimal length corrections through the modified background dynamics.

To describe the inflationary period, the slow-roll parameters are defined as follows
\begin{align}\label{eq:SRH}
\epsilon \equiv -\frac{\dot{H}}{H^{2}},\quad
\eta \equiv \frac{\dot{\epsilon}}{H\,\epsilon}.
\end{align}
The slow-roll inflation occurs provided that these parameters satisfy the conditions $(|\epsilon|, |\eta|) \ll 1$.
The evolutions of scalar and tensor perturbations are commonly described by their corresponding power spectra, ${\cal P}_{s}$ and ${\cal P}_{t}$. Under the slow-roll approximation, these spectra can be evaluated at the time of horizon crossing $k = aH$ as follows
\begin{eqnarray}
&\mathcal{P}_s\simeq\dfrac{H^2}{8 \pi ^{2}M_{\rm p}^{2} \epsilon}\Big|_{k=aH},\label{eq:Ps-SR} \\
&\mathcal{P}_t\simeq\dfrac{2H^2}{\pi ^{2}M_{\rm p}^2}\Big|_{k=aH} \label{eq:Pt-SR}.
\end{eqnarray}
The value of the scalar power spectrum at the pivot scale $k_\ast = 0.05~{\rm Mpc^{-1}}$ has been determined by Planck measurements as ${\cal P}_{s}(k_\ast) \simeq 2.1 \times 10^{-9}$ \cite{Planck:2018inflation,bk:18}.
Considering Eqs. (\ref{eq:Ps-SR}) and (\ref{eq:Pt-SR}), the scalar and tensor spectral indexes in Eq. (\ref{eq:nsnt}) can be estimated as
\begin{align}
& n_{s} \simeq 1+\left(1+\beta\,l_{\rm p}^{2}\, k^2\right)(-2\epsilon-\eta), \label{eq:ns_new_revised} \\
& n_{t} \simeq -\left(1+\beta\,l_{\rm p}^{2}\, k^2\right)(2\epsilon). \label{eq:nt_new_revised}
\end{align}
The latest constraint on the scalar spectral index imposed by the joint observational data from ACT DR6, Planck 2018, DESI BAO, and BICEP/Keck, is  $n_s = 0.974 \pm 0.003$~\cite{ACT:DR6params,ACT:DR6models}.
In addition, the tensor-to-scalar ratio $r$ is given by
\begin{align}
r\equiv \dfrac{\mathcal{P}_t}{\mathcal{P}_s} \simeq 16\epsilon. \label{eq:r_std_revised}
\end{align}
The observational constraint on $r$ from the latest data of Planck and BICEP/
Keck 2018 is $r < 0.036$.
In the single field inflationary models, the slow-roll conditions are satisfied when $\dot{\phi}^{2}/2 \ll V(\phi)$ and $|\ddot{\phi}|\ll |3H\dot{\phi}|$. There under,
the equation of motion (\ref{eq:KG}) simplifies to
\begin{equation}\label{eq:KGSR}
3H\dot{\phi }+V'(\phi ) \simeq 0,
\end{equation}
and the Hubble parameter $H$, derived from the modified Friedmann equation (\ref{eq:FR1-ML}), can be expressed approximately in terms of the potential $V(\phi)$ as follows
\begin{align}\label{eq:Happ}
H(\phi) \simeq \sqrt{\frac{V(\phi)}{3 M_{\rm p}^2}} \left(1 + \frac{\beta\,l_{\rm p}^2}{48 M_{\rm p}^2} V(\phi) \right).
\end{align}
It should be noted that in the limit $\beta=0$ (in the absence of minimal length effect), Eq.~(\ref{eq:Happ}) reduces to the standard slow-roll expression for the Hubble parameter, $H^2 \simeq V(\phi)/(3M_{\rm p}^2)$.

Considering the modified Friedmann equation (\ref{eq:Happ}), the first and second slow-roll parameters defined in Eq.~(\ref{eq:SRH}) can be approximated as
\begin{align}\label{eq:epSR_approx}
\epsilon \simeq \dfrac{3456\,M_{\rm p}^6 \big(16\,M_{\rm p}^2 + \beta\,l_{\rm p}^2\,V\big)\,V'^2}{V^2 \left(48 M_{\rm p}^2 +  \beta\, l_{\rm p}^2 V\right)^3},
\end{align}
%
{\small
\begin{align}\label{eq:etaSR_approx}
\eta \simeq \frac{
4608 M_{\rm p}^6 \Big[
2 \Big(384 M_{\rm p}^4 + \beta\,l_{\rm p}^2 V \left(32 M_{\rm p}^2 + \beta\,l_{\rm p}^2 V \right)\Big) V'^2
- V \left(16 M_{\rm p}^2 + \beta\,l_{\rm p}^2 V \right) \left(48 M_{\rm p}^2 + \beta\,l_{\rm p}^2 V \right) V''
\Big]
}{
V^2 \left(16 M_{\rm p}^2 + \beta\,l_{\rm p}^2 V \right) \left(48 M_{\rm p}^2 + \beta\,l_{\rm p}^2 V \right)^3
}.
\end{align}}
In the above expressions, $V$, $V'$, and $V''$ are functions of $\phi$, and we have dropped the argument for ease of notation.
Noteworthy, in the limit $\beta=0$, Eqs.~(\ref{eq:epSR_approx}) and (\ref{eq:etaSR_approx}) reduce to $\epsilon \approx \epsilon_V$ and $\eta \approx  4\epsilon_V-2\eta_V$, respectively. Here, $\epsilon_V \equiv \frac{M_{\rm p}^2}{2}\left(\frac{V'}{V}\right)^2$ and $\eta_V \equiv M_{\rm p}^2\left(\frac{V''}{V}\right)$ are the canonical potential slow-roll parameters.

In what follows, we consider the well known $\alpha$-attractor T-model potential for our model which is given by
\begin{align}\label{eq:V_alpha_revised}
V(\phi) = \Lambda^4 \tanh^2\left(\frac{\phi}{\sqrt{6\alpha}M_{\rm p}} \right),
\end{align}
where $\Lambda$ is the energy scale of the model with dimension of mass, and it can be defined through fixing the amplitude of scalar perturbations at the pivot scale. Furthermore, $\alpha$ is a dimensionless parameter characteristic of the model.
It is worth noting that we focus on the T-model potential as a representative example of the $\alpha$-attractor class. As our results will show, consistency with observational data in the presence of GUP corrections requires large values of the parameter $\beta$, which in turn favors the small-$\alpha$ regime. In this limit, the closely related E-models yield nearly identical predictions for $n_s$ and $r$. Therefore, an analysis of the E-model would not alter the main conclusions of this work. For clarity, we limit our scope to the T-model.

In the following section, the $\alpha$-attractor T-model potential will be examined within the GUP framework.
The primary objective is to analyze how minimal length effects influence the model predictions for key inflationary observables,  such as $n_s$ and $r$. This analysis will be carried out in light of recent observational data, particularly from the ACT DR6 release, to assess whether incorporating quantum gravity corrections improves the model consistency with these high precision measurements.

\section{Observational viability of the model}\label{sec:observational_viability}

Before directly confronting the predictions of the $\alpha$-attractor T-model modified by GUP with observational contours for $n_s$ and $r$, it is instructive to first examine the intrinsic behavior of the key dynamical quantities under the influence of the GUP corrections. This understanding will provide a foundation for interpreting the derived observational constraints.

\subsection{Behavior of background quantities with GUP corrections}
\label{subsec:background_dynamics}

\begin{figure}[H]
\centering
\includegraphics[width=0.6\textwidth]{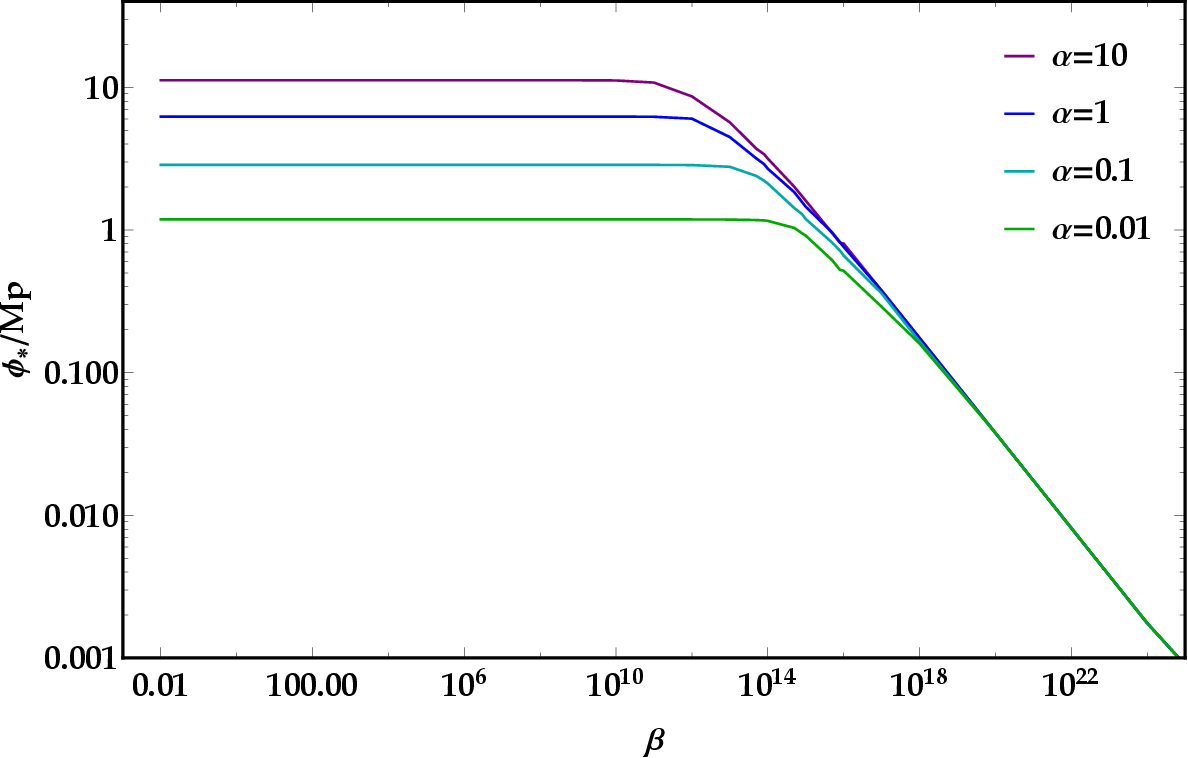}
\caption{Field value of the scalar field at the pivot scale $\phi_*/M_{\rm p}$ versus the GUP parameter $\beta$ for different values of $\alpha$. Here, $N_*=60$ is assumed for the horizon exit of the pivot scale.}
\label{fig:phi_star_vs_beta}
\end{figure}

We begin by analyzing the variation of the scalar field at the pivot scale $\phi_{*}$ as a function of GUP parameter $\beta$, as illustrated in Fig.~\ref{fig:phi_star_vs_beta}. This analysis considers four different values of $\alpha$.
Figure \ref{fig:phi_star_vs_beta} shows that
i) for a given $\beta$, the value of $\phi_*$ increases when the parameter $\alpha$ increases.
This indicates that the presence of a minimal length changes the initial value of the inflaton, which can have a significant impact on the model predictions.
ii) For a given $\alpha$, the value of $\phi_*$ decreases as the GUP parameter $\beta$ increases. This reduction is a key advantage of the model with GUP corrections. For large values of $\beta$, the inflaton field $\phi_*$ is driven into the sub-Planckian domain. This ensures the inflationary dynamics operate within a valid perturbative regime, which greatly enhances the theoretical self-consistency of the model.
iii) For $\beta \gtrsim 10^{18}$, the scalar field at the pivot scale $\phi_{*}$ becomes independent of the parameter $\alpha$.
%
Note that throughout the paper, we set $N_{*}=60$ for the horizon exit of the pivot scale.
Additionally, all parameters involving mass are normalized by the reduced Planck mass $M_{\rm p}$.

\begin{figure}[H]
\centering
\begin{minipage}[b]{1\textwidth}
\centering
\subfigure[\label{fig-phia1} ]{ \includegraphics[width=0.47\textwidth]%
{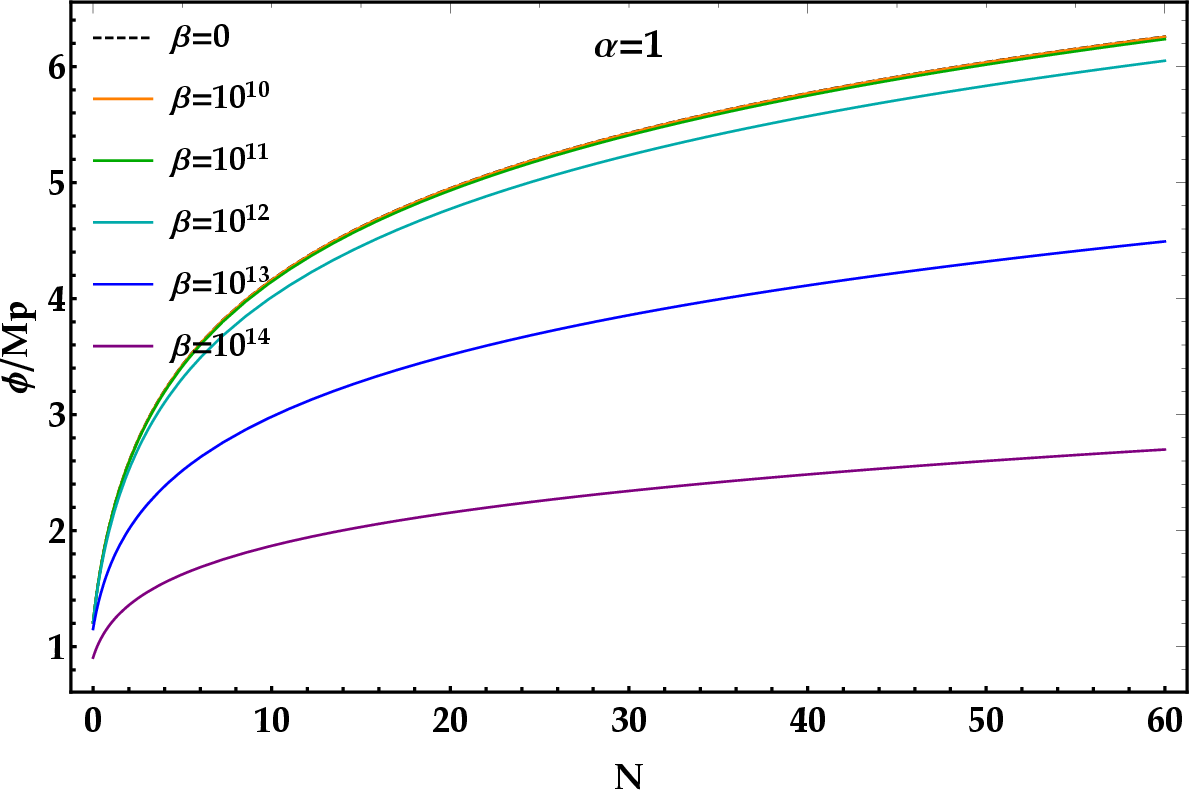}}
\hspace*{0.02\textwidth}
\subfigure[\label{fig-phia10} ]{ \includegraphics[width=0.47\textwidth]%
{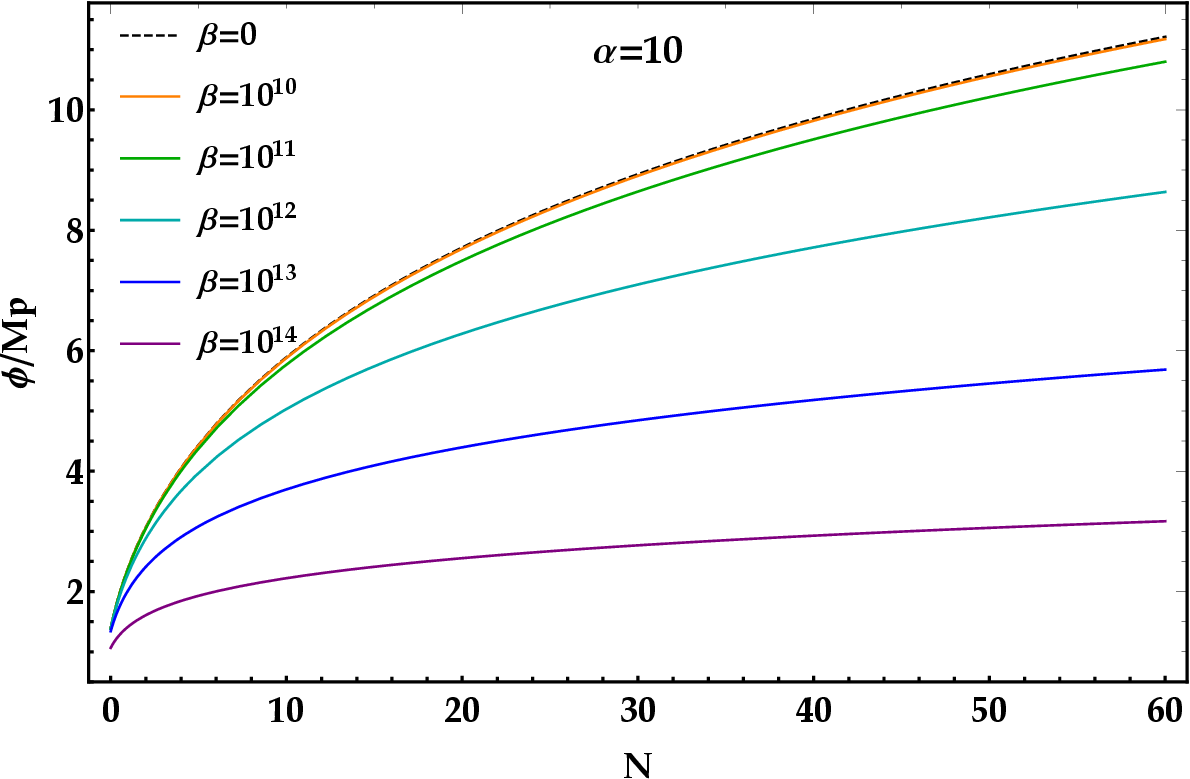}}
\end{minipage}
\caption{Evolution of the scalar field $\phi/M_p$ as a function of $e$-folds number $N$ for (a) $\alpha=1$ and (b) $\alpha=10$. Each panel displays six curves corresponding to different values of the GUP parameter $\beta=(0, 10^{10}, 10^{11}, 10^{12}, 10^{13}, 10^{14})$. The case of $\beta=0$ refers to the standard model of inflation.}
\label{fig:phi_evolution}
\end{figure}

Substituting Eqs. (\ref{eq:Happ}) and (\ref{eq:V_alpha_revised}) into the equation of motion (\ref{eq:KGSR}) and solving the resulting equation numerically, one can obtain the evolution of the scalar field $\phi$ as a function of the $e$-folds number $N$ as depicted in Fig.~\ref{fig:phi_evolution}.
In each panel of this figure, six distinct curves illustrate the impact of varying the GUP parameter $\beta$.
A general feature observed in both panels is that the scalar field $\phi$ decreases throughout inflation, which is consistent with the inflaton rolling down its potential.
Additionally, the parameter $\beta$ can significantly influence this evolution.
For fixed values of $\alpha$ and $N$, larger values of $\beta$ lead to smaller values of $\phi$.
The effect of GUP corrections becomes noticeable only when $\beta$ exceeds a certain threshold.
According to the figure, the scalar field values remain nearly unchanged for $\beta \lesssim \mathcal{O}(10^{11})$ when $\alpha = 1$, and for $\beta \lesssim \mathcal{O}(10^{10})$ when $\alpha = 10$. This behavior suggests that the model sensitivity to GUP corrections depends on the value of $\alpha$.
Note that the case $\beta = 0$ corresponds to the standard inflationary model, in which minimal length effects are absent.

\begin{figure}[H]
\centering
\begin{minipage}[b]{1\textwidth} 
\centering 
\subfigure[\label{fig-Ha1} ]{ \includegraphics[width=0.47\textwidth]%
{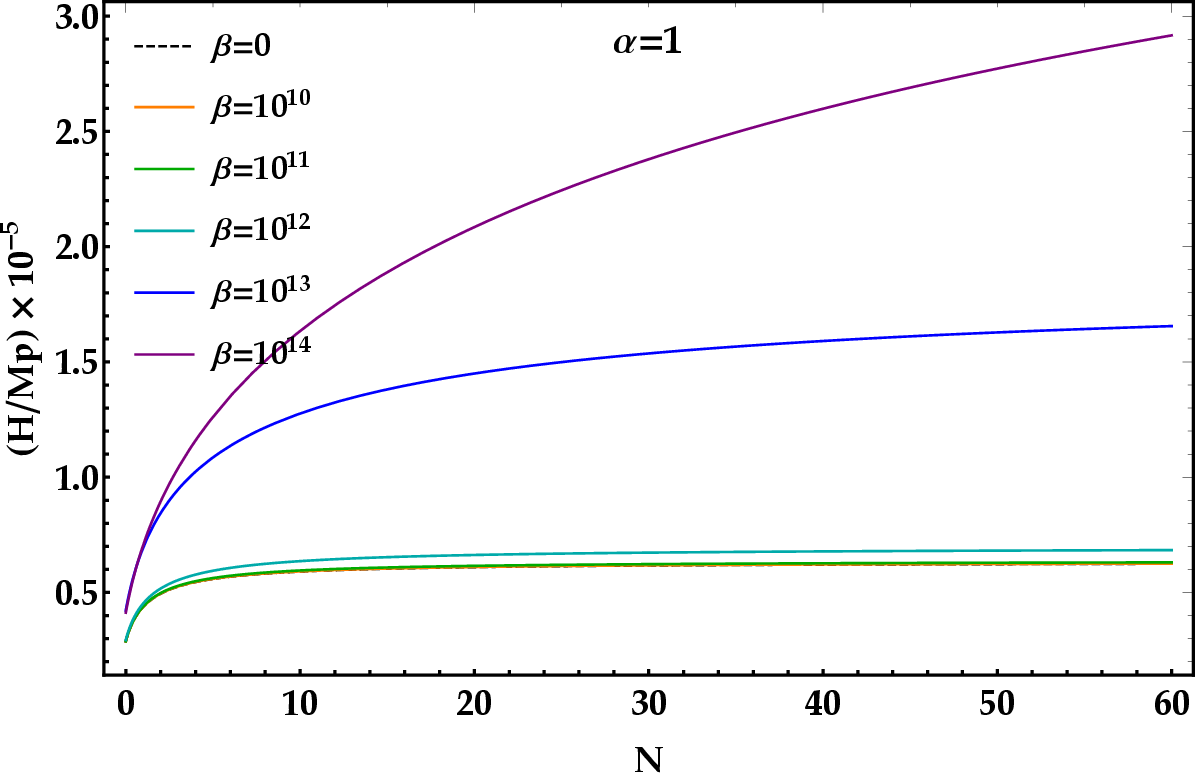}}
\hspace*{0.02\textwidth} 
\subfigure[\label{fig-Ha10} ]{ \includegraphics[width=0.47\textwidth]%
{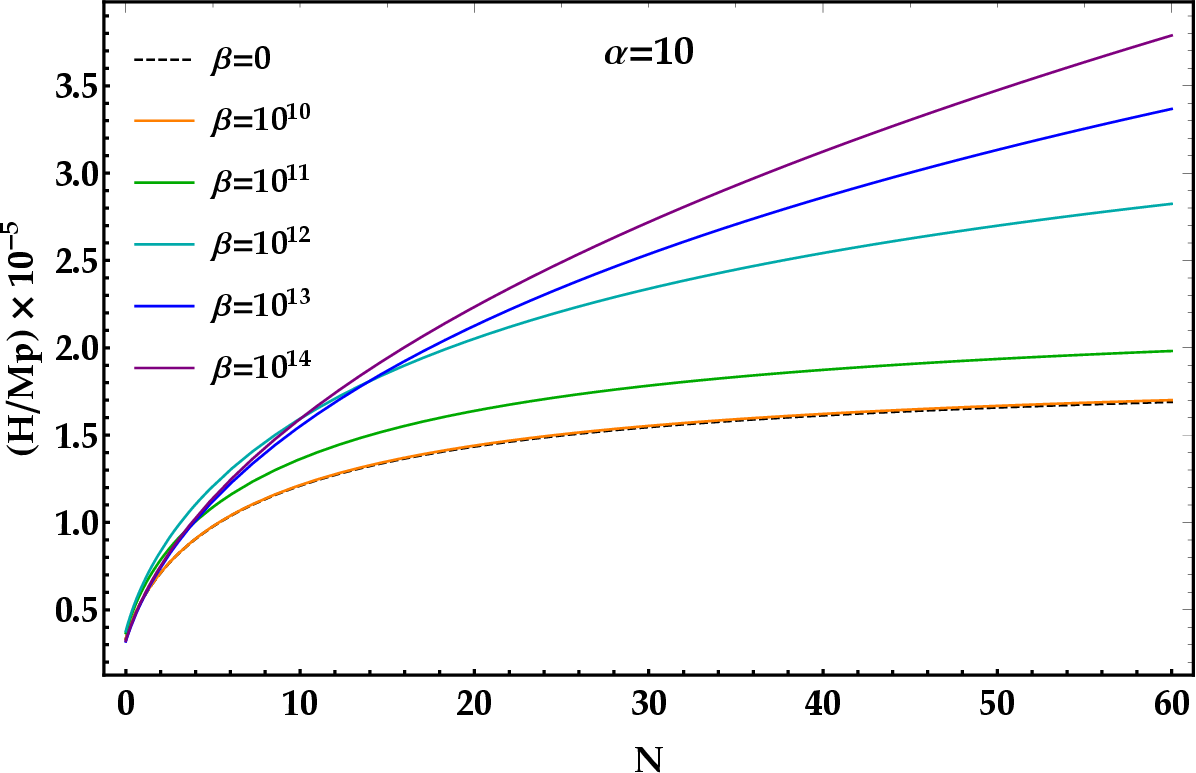}}
\end{minipage}
\caption{Evolution of the Hubble parameter $H$ as a function of $e$-folds number $N$ for (a) $\alpha=1$ and (b) $\alpha=10$. Each panel displays six curves corresponding to different values of the GUP parameter $\beta=(0, 10^{10}, 10^{11}, 10^{12}, 10^{13},10^{14})$.}
\label{fig:H_evolution}
\end{figure}

Complementing the evolution of the scalar field, Fig.~\ref{fig:H_evolution} illustrates the behavior of the Hubble parameter $H$, Eq. (\ref{eq:Happ}), as a function of the $e$-folds number $N$.
The left and right panels of Fig.~\ref{fig:H_evolution}  show the results for $\alpha=1$ and $\alpha=10$, respectively.
As in the previous figure, each panel contains six curves corresponding to different values of the GUP parameter $\beta$.
Figure \ref{fig:H_evolution} shows that the Hubble parameter decreases during inflation, which is a typical feature of slow-roll dynamics.
The presence of the GUP parameter $\beta$ modifies this evolution. For fixed values of $\alpha$ and $N$, increasing $\beta$ leads to a higher value of the Hubble parameter.
The detailed behavior of $H(N)$ plays a crucial role in determining the slow-roll parameters $\epsilon$ and $\eta$, which subsequently influence $n_s$ and $r$.
As the evolution of $H$ is altered by GUP corrections, it is expected that these modifications will affect the theoretical predictions of the model.
In the following analysis, the model predictions at large scales will be evaluated and compared with the most recent observational constraints deduced from ACT data.

\subsection{\texorpdfstring{Inflationary observables $n_s$ and $r$ in the GUP framework}{Inflationary observables n-s and r in the GUP framework}}
\begin{figure}[H]
\centering
\includegraphics[width=0.5\textwidth]{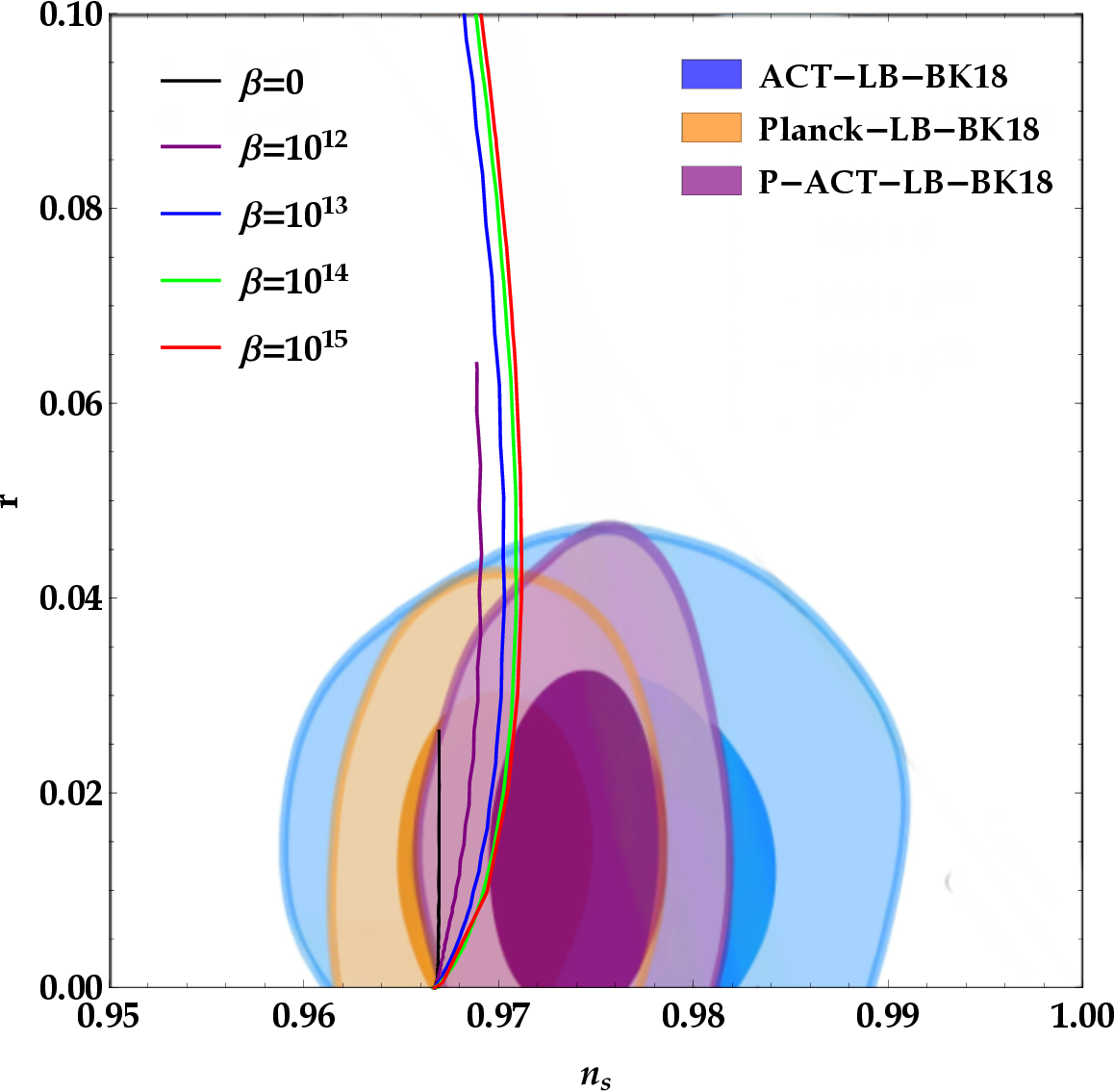} 
\caption{The $r-n_s$ plane showing predictions for the $\alpha$-attractor T-model modified by GUP. The $r-n_s$ curves correspond to fixed GUP parameter values: $\beta=0$ (black), $\beta=10^{12}$ (purple), $\beta=10^{13}$ (blue), $\beta=10^{14}$ (green), and $\beta=10^{15}$ (red). Along each curve, $\alpha$ varies from $0.001$ at the bottom to $10$ at the top.}
\label{fig:r_ns_beta_fixed}
\end{figure}

\begin{figure}[H]
\centering
\begin{minipage}[b]{1\textwidth} 
\centering 
\subfigure[\label{fig-nsB} ]{ \includegraphics[width=0.47\textwidth]%
{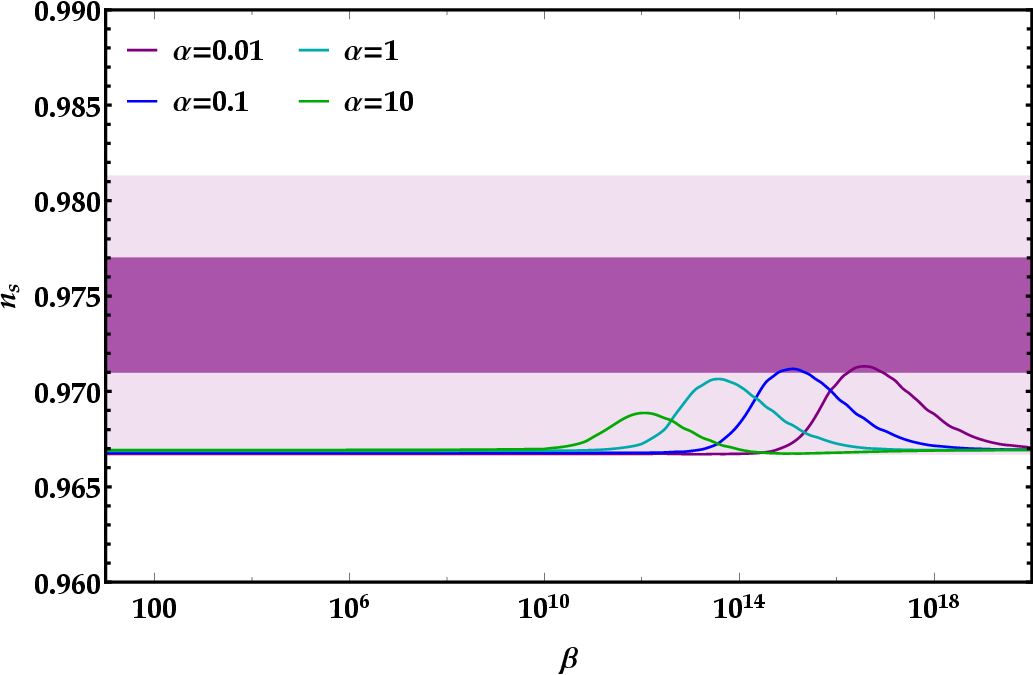}}
\hspace*{0.02\textwidth} 
\subfigure[\label{fig-rB} ]{ \includegraphics[width=0.47\textwidth]%
{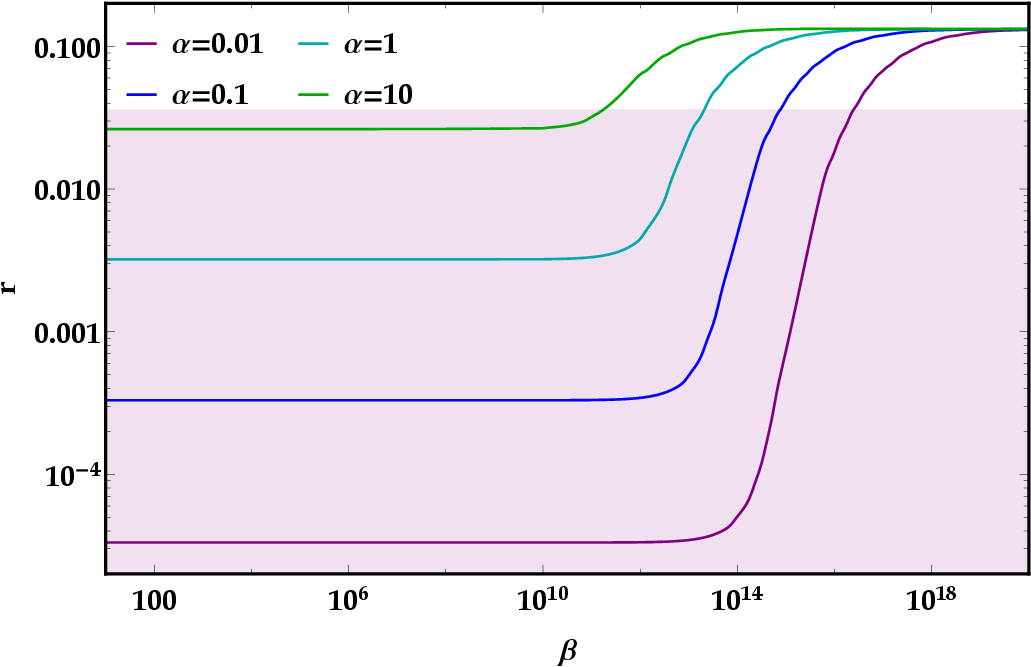}}
\end{minipage}
\caption{The inflationary observables as a function of the GUP parameter $\beta$ for several fixed values of $\alpha$. (a) The scalar spectral index $n_s$ versus $\beta$. (b) The tensor-to-scalar ratio $r$ versus $\beta$ on a logarithmic scale. The dark and light purple shaded regions show the 68\% and 95\% CL constraints from the combined P-ACT-LB-BK18 dataset, respectively.}
\label{fig:r_ns_alpha_fixed}
\end{figure}

\begin{figure}[H]
\centering
\includegraphics[width=0.5\textwidth]{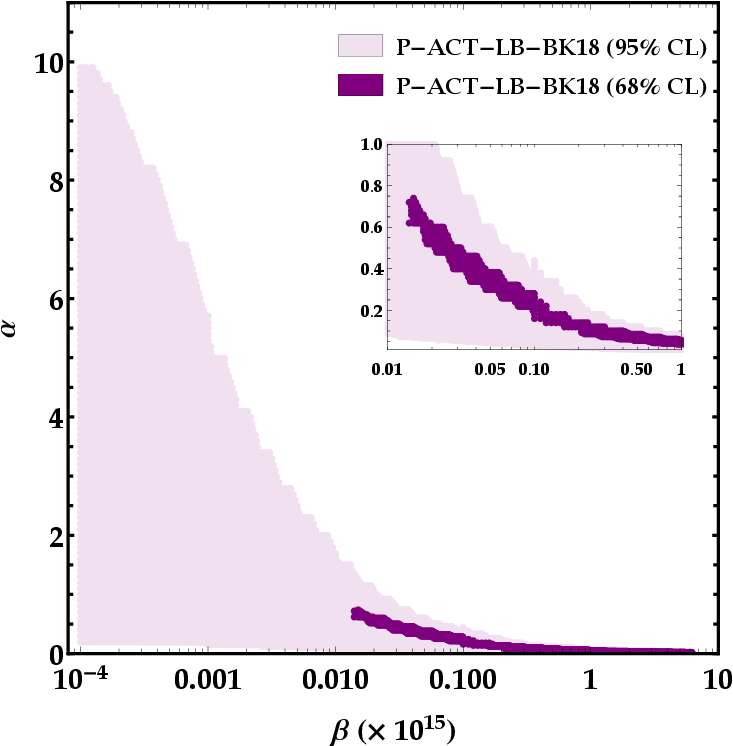} 
\caption{The phase-space diagram showing the allowed regions for the parameter $\alpha$ and the GUP parameter $\beta$ constrained by P-ACT-LB-BK18 data at 68\% CL (dark purple) and 95\% CL (light purple). The inset shows a magnified view of the region for smaller $\alpha$ values.}
\label{fig:phase_space_alpha_beta}
\end{figure}

Here, we compare the theoretical predictions of $\alpha$-attractor T-model modified by GUP with recent observational data to assess its viability.
Using Eqs. (\ref{eq:ns_new_revised}) and (\ref{eq:r_std_revised}), the model predictions in the $r-n_s$ plane are displayed in Fig. \ref{fig:r_ns_beta_fixed}, overlaid with the 68\% and 95\% CL contours from the combined P-ACT-LB-BK18 dataset, which includes ACT DR6, Planck, DESI BAO, and BICEP/Keck 2018 results~\cite{ACT:DR6params,ACT:DR6models}.
The plot features five distinct curves, each corresponds to a fixed value of the GUP parameter $\beta$. Along each curve, the parameter $\alpha$ varies from $0.001$ at the bottom to $10$ at the top. As shown in the figure, for $\beta = 0$, the predictions with $N_* = 60$ lie near the edge of the $2\sigma$ CL contour. However, as $\beta$ increases, the predicted values of $n_s$ and $r$ shift accordingly.
Notably, for $\beta \gtrsim 10^{13}$, the predictions of the model fall well within the 68\% CL region, indicating that the inclusion of GUP effects significantly enhances the agreement between theoretical predictions and observational data. This improvement is accompanied by a simultaneous increase in both $n_s$ and $r$, effectively shifting the predictions upward and to the right in the $r-n_s$ plane. Consequently, for higher values of $\beta$, smaller values of $\alpha$ are required for the model to remain consistent with the latest observational constraints.
To better understand the direct impact of the GUP parameter on the inflationary observables, we plot the scalar spectral index $n_s$ and the tensor-to-scalar ratio $r$ as a function of $\beta$ for different values of $\alpha$ (Fig. \ref{fig:r_ns_alpha_fixed}).
Figure \ref{fig-nsB} shows the behavior of $n_s$ versus $\beta$. The plot reveals a non-monotonic relationship: for each value of $\alpha$, as $\beta$ increases, $n_s$ first rises from its standard value, reaches a peak within the observationally favored region, and then decreases. This highlights that for any given $\alpha$, GUP corrections can successfully shift $n_s$ into the 68\% or 95\% CL regions, but only for a specific range of $\beta$ values.
Figure \ref{fig-rB} displays the tensor-to-scalar ratio $r$ as a function of $\beta$. In contrast to the scalar spectral index, $r$ shows a monotonic increase, rising sharply from a small value to a higher plateau as $\beta$ becomes large. This plot also indicates that as the required value of $\beta$ increases, smaller values of  $\alpha$ are necessary for the model to remain compatible with the observational bound on $r$.
To further investigate the viable parameter space, Fig.~\ref{fig:phase_space_alpha_beta} presents a phase-space diagram delineating the allowed regions in the $\alpha$–$\beta$ plane, based on the 68\% and 95\% CL from the combined P-ACT-LB-BK18 dataset. The corresponding numerical values extracted from the figure are summarized in Table~\ref{tab1}.
The analysis indicates that the maximum value of $\alpha$ for which the model predictions lie within the 68\% CL contour is approximately $\alpha \simeq 0.74$, which corresponds to the GUP parameter $\beta \simeq 1.4 \times 10^{13}$. Note that this value for $\beta$ is in well agreement with upper bounds on the GUP parameter deduced from cosmological analysis as well as quantum and gravitational experiments \cite{Luciano:2021}.
Moreover, the results demonstrate that for higher values of $\beta$, the model predictions remain consistent with the 68\% CL region only for comparatively lower values of $\alpha$. Table~\ref{tab1} further shows that as the magnitude of $\beta$ increases, the allowed range of $\alpha$ becomes increasingly narrow.
These findings strongly suggest that incorporating GUP effects can substantially improve the observational compatibility of $\alpha$-attractor inflationary model.



\begin{table}[htbp] 
  \centering
  \caption{Allowed ranges for the parameter $\alpha$ for different values of the GUP parameter $\beta$, based on 68\% CL and 95\% CL constraints from P-ACT-LB-BK18 data \cite{ACT:DR6models}.}
  \label{tab1}
  \begin{tabular}{ccc} 
    \hline\hline 
     $\beta$ & $\alpha ~(95\%$ CL) & $\alpha ~(68\% $CL) \\
    \hline 
     $[0,10^{11}]$ & $\alpha \leq 10$ & $-$ \\
     $ 10^{13}$ & $\alpha \leq 1.50$ & $-$ \\
     $ 1.4 \times 10^{13}$ & $\alpha \leq 1.35$ & $[0.62,0.74]$ \\
     $ 10^{14}$ & $\alpha \leq 0.43$ & $[0.15,0.23]$ \\
     $ 10^{15}$ & $\alpha \leq 0.07$ & $[0.037,0.054]$ \\
     $ 10^{16}$ & $\alpha \leq 0.02$ & $[0.008,0.011]$ \\
    \hline\hline 
  \end{tabular}
\end{table}

\section{Conclusion}\label{conc}
In this paper, the $\alpha$-attractor T-model of inflation was investigated, incorporating the effects of a minimal measurable length as implemented through the GUP. The study focused on assessing the observational viability of this GUP scenario against recent cosmological data, primarily from ACT DR6 combined with Planck, DESI BAO, and BICEP/Keck 2018 observations. The GUP corrections were introduced via modified Friedmann equations and altered definitions for the spectral indices.

The analysis of the inflationary background dynamics indicates that the GUP parameter $\beta$ has a significant impact on key cosmological quantities.
The evolution of the scalar field $\phi(N)$ and the Hubble parameter $H(N)$ during inflation is notably altered in the presence of GUP corrections.
An increase in $\beta$ leads to smaller values of the scalar field and a higher Hubble rate.
These corrections become significant only when $\beta$ exceeds a certain threshold, as demonstrated in Figs.~\ref{fig:phi_evolution} and \ref{fig:H_evolution}.
The analysis indicates that the model sensitivity to GUP effects depends on the value of $\alpha$.

We concluded that incorporating GUP corrections into the $\alpha$-attractor T-model leads to a markedly improved agreement with recent observational data.
In particular, while the standard scenario with $\beta = 0$ places the model predictions for $N_* = 60$ near the edge of the $2\sigma$ region, the presence of minimal length effects for $\beta \gtrsim 10^{13}$ shifts the predicted values of $n_s$ and $r$ into the 68\% confidence level region favored by the P-ACT-LB-BK18 dataset.
A key outcome of our analysis is that the GUP parameter $\beta$ directly impacts the inflationary observables. We found that increasing $\beta$ leads to a simultaneous increase in both the scalar spectral index $n_s$ and the tensor-to-scalar ratio $r$. This effect is crucial, as it shifts the model predictions into closer agreement with the latest observational data.

Notably, a maximum value of $\alpha \simeq 0.74$ remains compatible with the 68\% confidence region at $\beta \simeq 1.4 \times 10^{13}$.
Furthermore, smaller values of $\alpha$ are found to be viable at larger $\beta$, although the allowed interval for $\alpha$ becomes increasingly narrow as $\beta$ rises.
These findings underscore the potential of GUP-induced modifications to enhance the observational viability of $\alpha$-attractor model of inflation.

\subsection*{Acknowledgements}
The authors thank the referee for his/her valuable comments.


\begin{thebibliography}{99}
\bibitem{Starobinsky:1980} A. A. Starobinsky, Phys. Lett. B \textbf{91}, 99 (1980).
\bibitem{Guth:1981} A. H. Guth, Phys. Rev. D \textbf{23}, 347 (1981).
\bibitem{Linde:1982} A. D. Linde, Phys. Lett. B \textbf{108}, 389 (1982).
\bibitem{Mukhanov:1981} V. F. Mukhanov and G. V. Chibisov, J. Exp. Theor. Phys. lett. \textbf{33}, 532 (1981).
\bibitem{Hawking:1982} S. W. Hawking, Phys. Lett. B \textbf{115}, 295 (1982).
\bibitem{Guth:1982pi} A. H. Guth and S.-Y. Pi, Phys. Rev. Lett. \textbf{49}, 1110 (1982).
\bibitem{Bardeen:1983} J. M. Bardeen, P. J. Steinhardt and M. S. Turner, Phys. Rev. D \textbf{28}, 679 (1983).
\bibitem{Starobinsky:1982ee} A.~A.~Starobinsky, Phys. Lett. B \textbf{117}, 175 (1982).
\bibitem{WMAP:2013} G. Hinshaw et al. (WMAP Collaboration), Astrophys. J. Suppl. Ser. \textbf{208}, 19 (2013).
\bibitem{Planck:2018inflation} Y. Akrami et al. (Planck Collaboration), Astron. Astrophys. \textbf{641}, A9 (2020).
\bibitem{ACT:DR6params} T. Louis et al. (ACT Collaboration), arXiv:2503.14452.
\bibitem{ACT:DR6models} E. Calabrese et al. (ACT Collaboration), arXiv:2503.14454.
\bibitem{SPT:2022filters} L. Balkenhol et al. (SPT-3G Collaboration), Phys. Rev. D \textbf{108}, 023510 (2023).
\bibitem{Kallosh:2013a} R. Kallosh and A. Linde, J. Cosmol. Astropart. Phys. \textbf{07}, 002 (2013).
\bibitem{Ellis:2013} J. Ellis, D. V. Nanopoulos and K. A. Olive, J. Cosmol. Astropart. Phys. \textbf{10}, 009 (2013).
\bibitem{Kallosh:2013b} R. Kallosh, A. Linde and D. Roest, J. High Energy Phys. \textbf{11}, 198 (2013).
\bibitem{Ferrara:2013} S. Ferrara, R. Kallosh, A. Linde and M. Porrati, Phys. Rev. D \textbf{88}, 085038 (2013).
\bibitem{Kallosh:2014} R. Kallosh, A. Linde and D. Roest, Phys. Rev. Lett. \textbf{112}, 011303 (2014).
\bibitem{Carrasco:2015uma}
J. J. M. Carrasco, R. Kallosh, A. Linde and D. Roest,
Phys. Rev. D \textbf{92}, 041301 (2015).
\bibitem{Teimoori:2021} Z. Teimoori, K. Rezazadeh, M. A. Rasheed and K. Karami, J. Cosmol. Astropart. Phys. \textbf{10}, 018 (2021).
\bibitem{Rezazadeh:2022}
K. Rezazadeh, Z. Teimoori, S. Karimi and K. Karami, Eur. Phys. J. C \textbf{82}, 758 (2022).
\bibitem{bk:18}
P. A. R. Ade et al. (BICEP/Keck Collaboration), Phys. Rev. Lett. {\bf 127}, 151301 (2021).
\bibitem{Drees:2025}
M. Drees and Y. Xu, Phys. Lett. B \textbf{867}, 139612 (2025).
\bibitem{Dioguardi:2025mpp} C. Dioguardi and A. Karam, Phys. Rev. D \textbf{111}, 123521 (2025).
\bibitem{Dioguardi:2025} C. Dioguardi, A. J. Iovino and A. Racioppi, arXiv:2504.02809.
\bibitem{Wolf:2025ecy}
W. J. Wolf, arXiv:2506.12436.
\bibitem{Kallosh:2025act} R. Kallosh, A. Linde and D. Roest, arXiv:2503.21030.
\bibitem{Aoki:2025} S. Aoki, H. Otsuka and R. Yanagita, arXiv:2504.01622.
\bibitem{Berera:2025} A. Berera, S. Brahma, Z. Qiu, R. O. Ramos and G. S.Rodrigues, arXiv:2504.02655.
\bibitem{Raidal:2025} I. D. Gialamas, A. Karam, A. Racioppi and M. Raidal, arXiv:2504.06002.
\bibitem{Salvio:2025} A. Salvio, arXiv:2504.10488.
\bibitem{Antoniadis:2025} I. Antoniadis, J. Ellis, W. Ke, D. V. Nanopoulos and K. A. Olive, arXiv:2504.12283.
\bibitem{Kim:2025} J. Kim, X. Wang, Y.-l. Zhang and Z. Ren, arXiv:2504.12035.
\bibitem{Haque:2025} M. R. Haque, S. Pal and D. Paul, arXiv:2505.01517.
\bibitem{Peng:2025bws} Z. Z. Peng, Z. C. Chen and L. Liu, arXiv:2505.12816.
\bibitem{Liu:2025qca} L. Liu, Z. Yi and Y. Gong, arXiv:2505.02407.
\bibitem{Gialamas:2025ofz} I. D. Gialamas, T. Katsoulas and K. Tamvakis, arXiv:2505.03608.
\bibitem{He:2025bli} M. He, M. Hong and K. Mukaida, arXiv:2504.16069.
\bibitem{Konishi:1990} K. Konishi, G. Paffuti and P. Provero, Phys. Lett. B \textbf{234}, 276 (1990).
\bibitem{Amati:1989} D. Amati, M. Ciafaloni and G. Veneziano, Phys. Lett. B \textbf{216}, 41 (1989).
\bibitem{Gross:1988} D. J. Gross and P. F. Mende, Nucl. Phys. B \textbf{303}, 407 (1988).
\bibitem{Ashtekar:1997} A. Ashtekar and J. Lewandowski, Class. Quant. Grav. \textbf{14}, A55 (1997).
\bibitem{Rovelli:1995} C. Rovelli and L. Smolin, Nucl. Phys. B \textbf{442}, 593 (1995).
\bibitem{Reuter:2006} M. Reuter, J.M. Schwindt, J. High Energy Phys. \textbf{01}, 070 (2006).
\bibitem{Lauscher:2005} O. Lauscher and M. Reuter, J. High Energy Phys. \textbf{10}, 050 (2005).
\bibitem{Connes:2000} A. Connes, J. Math. Phys. \textbf{41}, 3832 (2000).
\bibitem{Seiberg:1999} N. Seiberg and E. Witten, J. High Energy Phys. \textbf{09}, 032 (1999).
\bibitem{Kempf:1995} A. Kempf, G. Mangano and R. B. Mann, Phys. Rev. D \textbf{52}, 1108 (1995).
\bibitem{Bosso:2023aht}
P.~Bosso, G.~G.~Luciano, L.~Petruzziello and F.~Wagner, Class. Quant. Grav. \textbf{40}, 195014 (2023).
\bibitem{Luo:2025}  S. S. Luo and Z. W. Feng, Eur. Phys. J. Plus \textbf{140}, 331 (2025).
\bibitem{Gialamas:2024ivq}
I.~D.~Gialamas, T.~J.~K{\"a}rkk{\"a}inen and L.~Marzola, Phys. Lett. B \textbf{856}, 138880 (2024).
\bibitem{Segreto:2024vtu}
S.~Segreto and G.~Montani, Eur. Phys. J. C \textbf{84}, 796 (2024).
\bibitem{Bitaj:2024} M. Bitaj, N. Rashidi, K. Nozari and M. Roushan, Phys. Dark Univ. \textbf{46}, 101552 (2024).
\bibitem{EPL:2023} N. Rashidi, M. Roushan and K. Nozari, Europhys. Lett. \textbf{142}, 39001 (2023).
\bibitem{Banerjee:2022slh}
I.~K.~Banerjee and U.~K.~Dey, Eur. Phys. J. C \textbf{83}, no.5, 428 (2023).
\bibitem{Barca:2023epu}
G.~Barca, G.~Montani and A.~Melchiorri, Phys. Rev. D \textbf{108}, no.6, 063505 (2023).
\bibitem{Du:2022} X. D. Du and C. Y. Long, J. High Energy Phys. \textbf{10}, 063 (2022).
\bibitem{Feng:2022} Z. W. Feng, X. Zhou and S. Q. Zhou, J. Cosmol. Astropart. Phys. \textbf{06}, 022 (2022).
\bibitem{Luciano:2021} G. G. Luciano, Eur. Phys. J. C \textbf{81}, 1086 (2021).
\bibitem{Feng:2020ams}
Z.~W.~Feng, G.~He, X.~Zhou, X.~L.~Mu and S.~Q.~Zhou, Eur. Phys. J. C \textbf{81}, no.8, 754 (2021).
\bibitem{Mohammadi:2017} A. Mohammadi, et. al., Annals Phys. \textbf{385}, 214 (2017).
\bibitem{Awad:2014} A. Awad and A. F. Ali, J. High Energy Phys. \textbf{06}, 093 (2014). 
\bibitem{Ali:2012} A. F. Ali, J. High Energy Phys. \textbf{09}, 067 (2012). 
\bibitem{Zhu:2009} T. Zhu, J. R. Ren and M. F. Li, Phys. Lett. B \textbf{674}, 204 (2009). 
\bibitem{Das:2008} S. Das and E. C. Vagenas, Phys. Rev. Lett. \textbf{101}, 221301 (2008). 
\bibitem{Bambi:2007ty}
C.~Bambi and F.~R.~Urban, Class. Quant. Grav. \textbf{25}, 095006 (2008).
\bibitem{Medved:2004} A. J. M. Medved and E. C. Vagenas, Phys. Rev. D \textbf{70}, 124021 (2004).
\end{thebibliography}
\end{document}